\documentclass[12pt]{iopart}

\usepackage{graphicx, epsfig, epstopdf, array, subfigure, bm, iopams}

\begin{document}

\title{Proximity breakdown of hydrides in superconducting niobium cavities}

\author{A Romanenko, F Barkov, L D Cooley and A Grassellino} 
\ead{aroman@fnal.gov}
\address{Fermi National Accelerator Laboratory, Batavia, IL 60510, USA}

\begin{abstract}
Many modern and proposed future particle accelerators rely on superconducting radio frequency cavities made of bulk niobium as primary particle accelerating structures. Such cavities suffer from the anomalous field dependence of their quality factors $Q_0$. High field degradation - so-called 'high field Q-slope' - is yet unexplained even though an empirical cure is known. Here we propose a mechanism based on the presence of proximity-coupled niobium hydrides, which can explain this effect. Furthermore, the same mechanism can be present in any surface-sensitive experiments or superconducting devices involving niobium.
\end{abstract}

\pacs{74.25.nn, 74.45.+c,  81.65.-b, 85.25.-j}

\maketitle
Superconducting niobium is an important technological material used for e.g. superconducting radio frequency (SRF) cavities and Josephson junctions. SRF cavities are primary particle accelerating structures in many modern accelerators (i.e. CEBAF, CESR, SNS) and the technology of choice for future ones (i.e. ILC, XFEL, FRIB, Project X). Bulk niobium is predominantly used to make such cavities due to its highest $T_c$ among pure elements and high enough critical fields. Quality factor Q$_0$ of SRF cavities is determined by the surface resistance via the formula:
\begin{equation}
Q_0 = \frac{\omega U}{P_\mathrm{diss}} = \frac{\omega \mu_0 \int_V H^2 dV}{\int_A R_\mathrm{s}(H) H^2 dA}
\end{equation}
where $\omega$ is the angular frequency, $U$ is stored energy calculated from the integral of the field over cavity volume, and $P_\mathrm{diss}$ is the power dissipated in cavity walls calculated as an integral of surface resistance over cavity walls. Extremely low surface resistance in superconducting state and hence very high quality factors $Q_0 > 10^{11}$ are achievable in such structures.

Several decades of research demonstrated that the microwave surface resistance of niobium in superconducting state is very sensitive to heat and surface chemical treatments. Such surface resistance is completely determined by the nanostructure within the magnetic field penetration depth $\lesssim$ 100~nanometers where the screening currents flow. In addition to the intrinsic BCS-described component coming from thermally excited quasiparticles there is a number of extrinsic contributions to losses. Some of such contributions are understood (trapped magnetic flux, field emission, multipacting), while many are not. Among the unknown mechanisms are the ones leading to the so-called \textit{low, medium, and high field Q-slopes}, which correspond to, respectively, a decrease of $R_\mathrm{s}$ in the field range of about 0-20~mT, mild increase in the range of 20-100~mT, and a drastic increase above about 100~mT (see~\cite{Hasan_book2} for review). For reaching highest accelerating gradients (highest sustainable surface fields) and for highest quality factors (lowest surface resistance) medium and high field Q-slopes present a significant obstacle. An empirical recipe has been developed to achieve highest gradients, which allows to eliminate the high field Q-slope via a combination of electropolishing and 90-145$^\circ$C baking in vacuum frequently referred to as ``mild baking''. No controlled way to mitigate the medium field Q-slope is known. Understanding the mechanisms behind these effects is of uttermost importance for further progress in the SRF field from both scientific and practical points of view.

In this article we propose that the high field Q-slope (HFQS) can be explained by the breakdown of the proximity effect acting on niobium hydrides in the near-surface layer. Furthermore, Q-degradation phenomenon due to large niobium hydrides (\emph{Q-disease})~\cite{Aune_Q_Disease_1990, Bonin_SRF_91, Antoine_SRF91_Hydrogen, Halbritter_Q_Disease_SRF_1993, Knobloch_Q_Disease_Overview} can be quantitatively described by the same model.

Niobium is a material which exhibits high affinity for hydrogen. Hydrogen can enter niobium whenever the natural oxide layer of niobium pentoxide Nb$_2$O$_5$ is absent. In niobium cavities it happens during standard chemical treatments (buffered chemical polishing and electropolishing) and during the cooldown and air exposure following high temperature furnace treatments~\cite{Faber_H_reabsorption}. Inside niobium the distribution of hydrogen is non-uniform and exhibits the near-surface segregation. High values of 1-20 at.\%~\cite{Antoine_SRF91_Hydrogen, Tajima_SRF03_Hydrogen, Romanenko_SUST_ERD_2011} were reported for the concentration of hydrogen in the first several tens of nanometers below the surface while bulk hydrogen content is typically in the $\lesssim$ 100 ppm range.  Vacuum heat treatments at T$=$600-900$^\circ$C decrease the bulk concentration but the near-surface region is weakly affected and very high hydrogen concentration remains. In low RRR ($\lesssim 50$) niobium most of hydrogen (bulk and surface) is trapped by interstitial impurities such as e.g. oxygen as manifested in the absence of Q-disease in cavities made of such material. In the case of high RRR ($\sim$300) cavities it is not the case and hydrogen can move and precipitate.

\begin{figure}[htb]
\includegraphics[width=\linewidth]{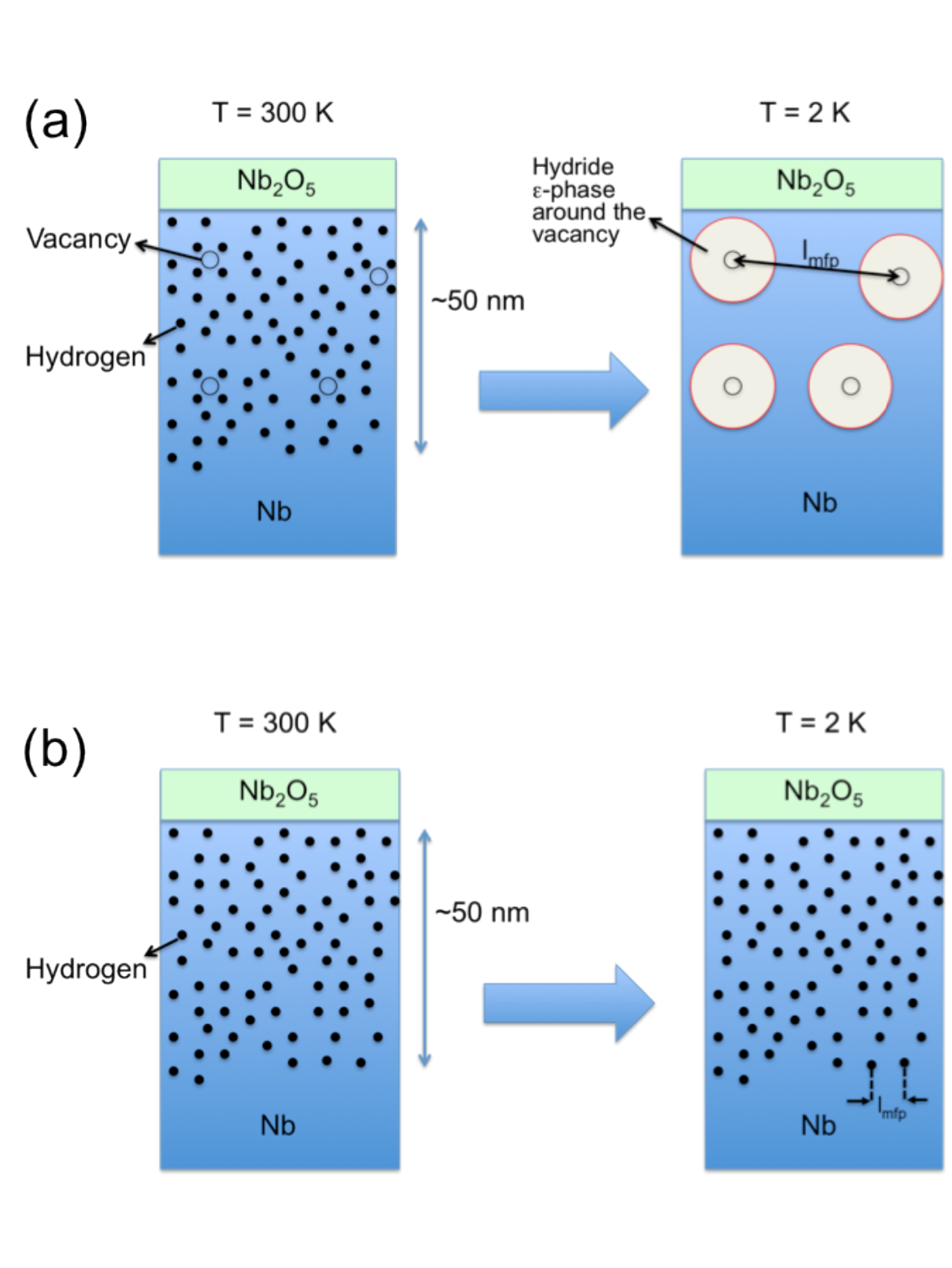}
\caption{\label{fig:Cooldown_schematic}Schematic of the effect of the cooldown on: (a) unbaked; (b) baked at 120$^\circ$C niobium. Precipitation of niobium hydride around nucleation centers (possibly vacancies) leads to the \textit{increase} of the electron m.f.p.}
\end{figure}

From NbH$_\mathrm{x}$ phase diagram~\cite{NbH_phases} it follows that the near-surface hydrogen-rich layer should transition from solid solution into $\epsilon$-phase of niobium hydride upon cooldown below $\sim$100-150~K depending on the concentration. Two important conditions which facilitate this transition are long enough diffusion length of hydrogen during the cooldown time in the precipitation temperature range and the presence of nucleation centers. Typically, a ``fast'' cooldown is performed on cavities to avoid the Q-disease, and it was believed that it allows to avoid any hydride precipitation. Nevertheless, a simple estimate based on the typical cooldown between 150 and 2~K (FNAL data was used) and the diffusion constant of hydrogen leads to the diffusion length of $L_\mathrm{diff} \sim 30~\mu \mathrm{m} \gg 50$~nm. This means that all the H within the segregation layer near the surface can precipitate during fast cooldown. Such a scenario is depicted in Fig.~\ref{fig:Cooldown_schematic}(a). Such a precipitation would lead to what is called a Q-disease only if there is a significant bulk hydrogen concentration and longer cooldown allowing hydrides to grow large.

Niobium hydrides were demonstrated to be normal conducting at T$>$1.5~K~\cite{NbH_Supercond_Prop_1965}. However, hydride precipitates will be subject to the superconducting proximity effect~\cite{DeGennes_Proximity_1965} and will have a non-zero order parameter up to a certain breakdown field $H_\mathrm{b}$ at which Meissner effect disappears. For the proximity effect in the normal/superconducting sandwich an important length scale is set by the normal metal coherence length \mbox{$\xi_\mathrm{N} = \hbar v_\mathrm{F} / (2\pi k_\mathrm{B} T)$}. For \mbox{T = 2~K} and assuming the same Fermi velocity $v_\mathrm{F}$ for hydrides as Nb ($1.4 \times 10^8$~cm/s) we get $\xi_\mathrm{N}\approx 800$~nm. Proximity effect does depend on temperature and two asymptotic cases $T \ll T_\mathrm{A}$ and $T \gg T_\mathrm{A}$ (where $T_A=\hbar v_\mathrm{F}/(2\pi d)$, $d$ is a normal metal thickness) were considered in the literature~\cite{PRB_Fauchere_Proximity_1997}. Based on the observations of the surface relief~\cite{Barkov_PRST_Hydrides_2012} the thickness $d$ of hydrides corresponding to the Q-disease is of order \mbox{100~nm $\ll \xi_\mathrm{N}$} and the corresponding $T_\mathrm{A} \approx 166~\mathrm{K} \gg 2~\mathrm{K}$. Hydrides should be even smaller in size for the high field Q-slope and hence in both cases we are in the limit of $T \ll T_\mathrm{A}$ and the breakdown field is provided by:
\begin{equation}
\label{eq:Hb}
H_\mathrm{b} \approx \frac{1}{6}\frac{\Phi_0}{\lambda_\mathrm{N} d}
\end{equation}
where $\Phi_0$ is the magnetic flux quantum, and \mbox{$\lambda_\mathrm{N}=\sqrt{mc^2/(4\pi n e^2)}$}. Here $n$ is the total electron density in the hydride, which may be different from that in niobium.

A direct evidence for the proximity effect in superconducting niobium cavities is provided by ``hydrogen Q-disease''. In this effect the microwave surface resistance does not change significantly up to peak surface magnetic fields of $\lesssim$10~mT followed by a sharp increase.  Such a threshold may be interpreted as a breakdown field of surface hydrides of the corresponding size. Surface relief measurements~\cite{Barkov_PRST_Hydrides_2012} of the hydrides formed under similar cooldown conditions to the ones leading to Q-disease show that the thickness of hydrides responsible for such Q-disease with $H_\mathrm{b}=10$~mT is $d \gtrsim 100$~nm. Then from Eq.~\ref{eq:Hb} it follows that the hydrides with average breakdown field of $\sim$100~mT corresponding to the HFQS onset should have smallest dimension of $d \gtrsim 10$~nm.

The proximity effect will also result in the suppression of the superconducting gap of the host niobium. For example, such a suppression attributed to NbO$_\mathrm{x}$ inclusions as a subject to the proximity effect was reported in the literature~\cite{Schwarz_Halbritter_JAP_1977, Halbritter_SRF_2001}. 

\begin{figure}[htb]
\includegraphics[width=\linewidth]{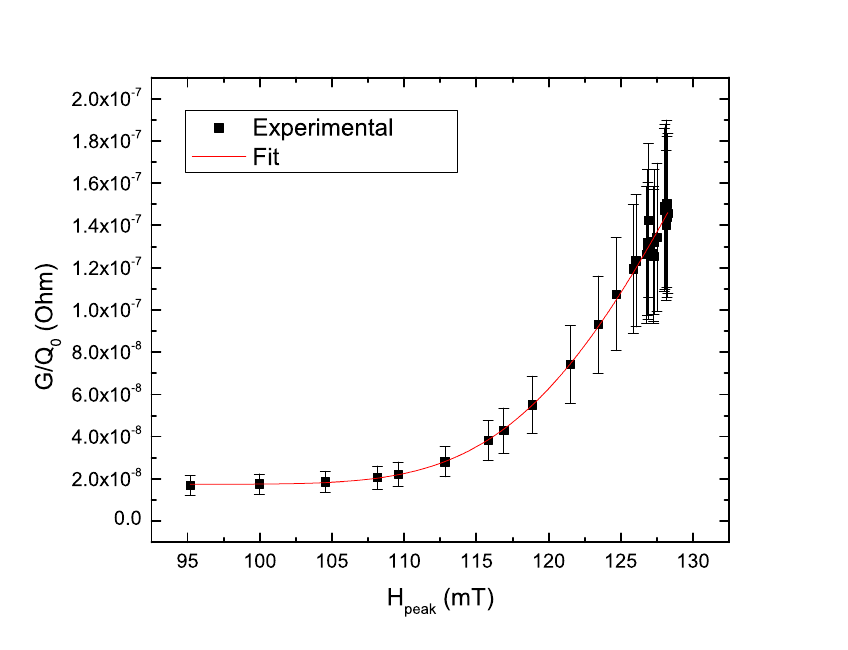}
\caption{\label{fig:Q_slope_fit}Fit of the high field Q-slope by Eq.~\ref{eq:Rs}. Best fit parameters: adjusted \mbox{$r^2=0.998$}, \mbox{$R_0=17.4$~n$\Omega$}, \mbox{$\alpha_s R_n=470$~n$\Omega$}, \mbox{$H_0=136$~mT}, \mbox{$\sigma=7.6\cdot10^{-4}$~mT$^{-1}$}.}
\end{figure}

\begin{figure}[htb]
\includegraphics[width=\linewidth]{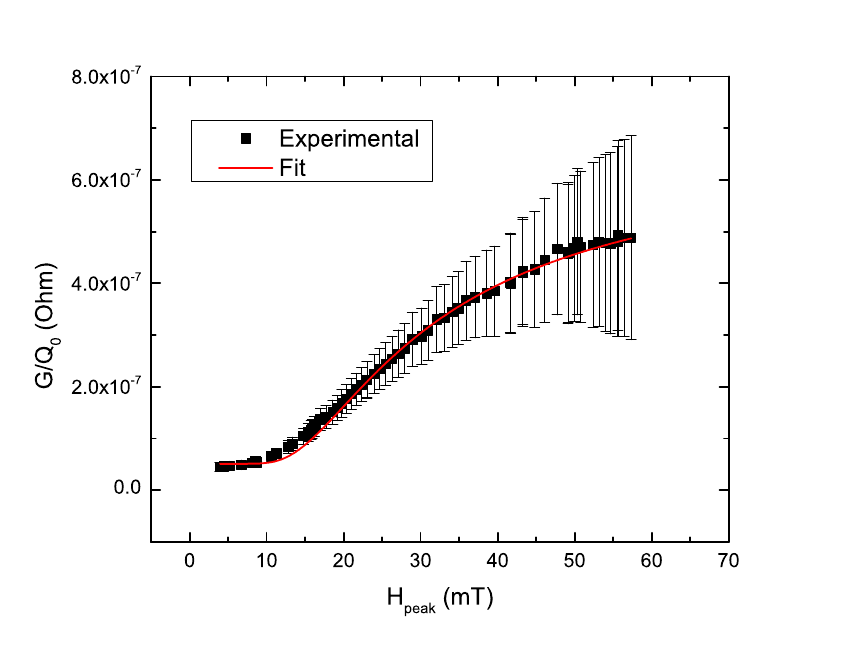}
\caption{\label{fig:Q_disease_fit}Fit of the Q-disease by Eq.~\ref{eq:Rs}. Best fit parameters: adjusted $r^2=0.997$, \mbox{$R_0=52$~n$\Omega$}, \mbox{$\alpha_s R_n=840$~n$\Omega$}, \mbox{$H_0=54$~mT}, \mbox{$\sigma=0.0281$~mT$^{-1}$}.}
\end{figure}

If we assume that hydrides are formed with the normal distribution of characteristic sizes $d$ then it follows from Eq.~\ref{eq:Hb} that the distribution of the breakdown fields $H_\mathrm{b}$ is determined by the distribution of $1/d$ and has the c.d.f.
\begin{equation}
F(H) = P\{H_\mathrm{b} < H\} = \frac{1}{2}\left[1-\mathrm{erf}\left(\frac{\frac{1}{H}-\frac{1}{H_0}}{\sqrt{2\sigma^2}}\right)\right]
\end{equation}
where $H_0$ is the mean and $\sigma$ is the width of the breakdown field distribution. If we attribute the normal conducting surface resistance $R_n$  and surface density $\alpha_s$ to niobium hydrides then the total surface resistance can be written as:
\begin{equation}
\label{eq:Rs}
R_\mathrm{s}(H) \approx R_0 + \alpha_\mathrm{s} R_\mathrm{n} \cdot F(H)
\end{equation}
Such surface resistance dependence on field provides an excellent fit to experimental data on cavities with the high field Q-slope and Q-disease available to us. Examples of fits using Eq.~\ref{eq:Rs} to the average surface resistance data are shown in Fig.~\ref{fig:Q_slope_fit} and Fig.~\ref{fig:Q_disease_fit} (average surface resistance is defined as $<R_\mathrm{s}>=G/Q_0$, where G is a constant determined by cavity geometry) .


As mentioned above, mild baking removes the high field Q-slope in all but fine grain BCP cavities. Based on the proposed mechanism of the high field Q-slope we suggest that the main effect of mild baking is the complete absence or significantly smaller in size hydrides. 

One of the effects of the mild baking on niobium is the drastic decrease in the electron m.f.p. within the penetration depth as deduced from fits to R$_s$(T) and directly measured lately with muon spin spectroscopy~\cite{Grassellino_TFSRF_2012}. Unlike other suggested mechanisms our model can explain m.f.p. change without involving diffusion of other interstitials (such as O, C, N) by purely the effect of the precipitation of hydrides . Indeed, if segregated hydrogen stays dispersed in the lattice then m.f.p. is determined by the average H-H distance. If the dispersed H precipitates into hydrides then the m.f.p. is either the mean hydride-hydride distance or determined by other less-abundant impurities (O, C, N), which in both cases is much larger than H-H distance before precipitation. The schematic of the near-surface structure of baked/unbaked cavities within our model is shown in Fig.~\ref{fig:Cooldown_schematic}.

Now we describe a possible mechanism of mild baking on the material level in further detail. It was demonstrated~\cite{Cizek_H_Defects_2005, Cizek_PRB_2009_Vac_H} that in the presence of a significant amount of hydrogen the formation energy of vacancies in niobium becomes much lower. This can lead to the very high concentrations of complexes consisting of a monovacancy and several hydrogen atoms (Vac-H). This process - a so-called ``superabundant vacancy'' formation~\cite{Fukai_PRL_1994_Pd_SAV} - was discovered in many metals including niobium. Here we propose that the near-surface layer enriched with hydrogen in SRF niobium cavities contains superabundant vacancies (SAV) after chemical (BCP or EP) or vacuum heat (600-800$^\circ$C) treatments. The concentration of SAVs can be estimated from~\cite{Cizek_H_Defects_2005}, which for hydrogen concentration \mbox{n$_\mathrm{H}$=10 at.\%} gives $\sim 10^{-3}$ at. \%.

It was discovered by positron annihilation studies~\cite{Alekseeva_PhysicaScripta_PAS_1979, Hautojarvi_Vac_H_PRB_1985} that Vac-H complexes start dissociating at T$\approx$90$^\circ$C, which agrees remarkably with the lowest effective temperature of mild baking found empirically. Lately direct positron annihilation Doppler broadening studies~\cite{Romanenko_TFSRF_2012} on cutout samples from baked and unbaked cavities showed a strong decrease in the vacancy concentration brought about by 120$^\circ$C bake. We suggest that Vac-H complexes dissociate during mild baking and the near-surface layer gets ``cleaned'' from vacancies by their diffusion to the surface and annihilation. 

Upon cooldown of SRF cavities to cryogenic temperatures, Vac-H complexes may serve as nucleation centers around which an $\epsilon$-phase of niobium hydride can form. In the case of the 120$^\circ$C baked niobium the nucleation centers are absent and hydrides do not form. Schematically such mechanism is depicted in Fig.~\ref{fig:Cooldown_schematic}. 

The distribution of precipitate sizes is determined by the kinetics of hydrogen diffusion towards nucleation centers during cooldown. If for simplicity we assume that hydrides are spherical then we can estimate their diameters. Taking Vac-H complexes as nucleation centers with concentration $c_\mathrm{vac} \sim 10^{-3}$ at.~\% and $c_\mathrm{H} \sim 5$ at.\% in the first 20~nm, assuming all H will precipitate from the layer determined by diffusion length over ``fast'' cooldown $L_\mathrm{diff} \sim 30~\mu \mathrm{m}$, and $c_H \sim 2$~ppm in the bulk we get $d \sim 6$~nm in a very good agreement with the estimates above based on the breakdown field if we assume that the breakdown field of spherical precipitates can be approximated by Eq.~\ref{eq:Hb} as well.

We suggest that if no Vac-H complexes are present, the NbH$_\mathrm{x}$ solution can be quenched to 2~K without any significant hydride formation, and no high field Q-slope is observed.

A number of testable predictions can be made based on our model. 

First, the $R_\mathrm{s}(H)$ dependence in Eq.~\ref{eq:Rs} predicts that $R_\mathrm{s}$ should change the character/saturate deep in the high field Q-slope when most of the hydrides become normal conducting. Such saturation is observed for hydrogen Q-disease and can be accessible at higher RF powers for the HFQS as well. It should be possible to check it with single cell cavities. We have observed hints of such saturation on two different cavities, but further testing at higher powers is required to confirm it. 

Second, if Vac-H mechanism for mild baking is correct then reintroduction of vacancies in the near-surface layer should lead to the reappearance of the high field Q-slope in the mild baked cavity. We are currently planning such an experiment with controlled introduction of vacancies by irradiation.

Finally, nanoscale hydrides may be directly observable using techniques such as an in situ cold stage imaging with TEM/STEM. Such a search is currently underway.

It should be noted further that if nanohydrides are present then the medium field Q-slope can be explained by the field dependence of the proximity effect below H$_b$. Hence it may open up a possibility for further significant improvement in niobium cavities. The problem may be approached by the combination of heat treatment and ``sealing'' the surface with the niobium pentoxide before reabsorption rebuilds the interstitial hydrogen content. Such experiments are currently underway at FNAL.

In summary, we have put forward a novel model for the high field Q-slope based on the proximity effect acting upon niobium hydrides within the penetration depth. Our model can also explain quantitatively the hydrogen Q-disease, which we suggest has the same physical mechanism. We propose that the mild baking effect is due to the strong suppression of hydride precipitation, which may be caused by the change in concentration of Vac-H complexes.

Authors would like to acknowledge useful discussions with H. Padamsee from Cornell. Fermilab is operated by Fermi Research Alliance, LLC under Contract No. De-AC02-07CH11359 with the United States Department of Energy. One of the authors (A. R.) acknowledges support under US DOE Office of Nuclear Physics Early Career Award for this work.

\section*{References}
\bibliographystyle{iopart-num}

\begin{thebibliography}{10}
\expandafter\ifx\csname url\endcsname\relax
  \def\url#1{{\tt #1}}\fi
\expandafter\ifx\csname urlprefix\endcsname\relax\def\urlprefix{URL }\fi
\providecommand{\eprint}[2][]{\url{#2}}

\bibitem{Hasan_book2}
Padamsee H 2009 {\em RF Superconductivity: Volume II: Science, Technology and
  Applications\/} (Wiley-VCH)

\bibitem{Aune_Q_Disease_1990}
Aune B, Bonin B, Cavedon J~M, Juillard M, Godin A, Henriot C, Leconte P, Safa
  H, Veyssiere A and Zylberajch C 1990 Degradation of niobium superconducting
  rf cavities during cooling times {\em Proceedings of the 1990 LINAC
  Conference\/} pp 253--255

\bibitem{Bonin_SRF_91}
Bonin B and Roth R~W 1991 Q degradation of niobium cavities due to hydrogen
  contamination {\em Proceedings of the 5th Workshop on RF Superconductivity\/}
  pp 210--244

\bibitem{Antoine_SRF91_Hydrogen}
Antoine C~Z, Aune B, Bonin B, Cavedon J, Juillard M, Godin A, Henriot C,
  Leconte P, Safa H, Veyssiere A, Chevarier A and Roux B 1991 The role of
  atomic hydrogen in {Q}-degradation of niobium superconducting rf cavities:
  analytical point of view {\em Proceedings of the Fifth Workshop on RF
  Superconductivity, DESY, Hamburg, Germany\/} pp 616--634

\bibitem{Halbritter_Q_Disease_SRF_1993}
Halbritter J, Kneisel P and Saito K 1993 {\em {Proceedings of the 6th Workshop
  on RF Superconductivity}\/} pp 617--627

\bibitem{Knobloch_Q_Disease_Overview}
Knobloch J 2003 {\em AIP Conference Proceedings\/} {\bf 671} 133--150

\bibitem{Faber_H_reabsorption}
Faber K and Schultze H 1972 {\em Scripta Metallurgica\/} {\bf 1} 1065--1070

\bibitem{Tajima_SRF03_Hydrogen}
Tajima T, Edwards R~L, Krawczyk F~L, Liu J, Schrage D~L, Shapiro A~H, Tesmer
  J~R, Wetteland C~J and Geng R~L 2003 {\em Proceedings of the 11th Workshop on
  RF Superconductivity\/} THP19

\bibitem{Romanenko_SUST_ERD_2011}
Romanenko A and Goncharova L~V 2011 {\em Supercond. Sci. Tech.\/} {\bf 24}
  105017

\bibitem{NbH_phases}
Manchester F~D and Pitre J~M 2000 {\em Phase Diagrams of Binary Hydrogen
  Alloys\/} (ASM International)

\bibitem{NbH_Supercond_Prop_1965}
Rauch G~C, Rose R~M and Wulff J 1965 {\em J. Less. Comm. Metals\/} {\bf 8}
  99--113

\bibitem{DeGennes_Proximity_1965}
Gennes P~G~D and Hurault J~P 1965 {\em Phys. Lett.\/} {\bf 17} 181

\bibitem{PRB_Fauchere_Proximity_1997}
Fauchere A and Blatter G 1997 {\em Phys. Rev. B\/} {\bf 56} 14102

\bibitem{Barkov_PRST_Hydrides_2012}
Barkov F, Romanenko A and Grassellino A 2012 {\em Phys. Rev. ST Accel. Beams\/}
  {\bf 15} 122001

\bibitem{Schwarz_Halbritter_JAP_1977}
Schwarz W and Halbritter J 1977 {\em J. Appl. Phys.\/} {\bf 48} 4618--4626

\bibitem{Halbritter_SRF_2001}
Halbritter J 2001 Material science of {Nb} {RF} accelerator cavities: where do
  we stand 2001? {\em Proceedings of the 10th Workshop on RF
  Superconductivity\/} (Tsukuba, Japan)

\bibitem{Grassellino_TFSRF_2012}
Grassellino A, Romanenko A, Barkov F and Suter A Near surface superconductivity
  of niobium cavity cutouts probed by low energy muon spin rotation, talk at
  {TFSRF'2012}, www.jlab.org/indico/conferenceDisplay.py?confId=22

\bibitem{Cizek_H_Defects_2005}
Cizek J, Prochazka I, Kuzel R, Becvar F, Cieslar M, Brauer G, Anwand W,
  Kirchheim R and Pundt A 2005 {\em J. of Alloys and Compounds\/} {\bf
  404--406} 580--583

\bibitem{Cizek_PRB_2009_Vac_H}
Cizek J, Prochazka I, Danis S, Brauer G, Anwand W, Gemma R, Nikitin E,
  Kirchheim R, Pundt A and Islamgaliev R 2009 {\em Phys. Rev. B\/} {\bf 79}
  054108

\bibitem{Fukai_PRL_1994_Pd_SAV}
Fukai Y and Okuma N 1994 {\em Phys. Rev. Lett.\/} {\bf 73} 1640--1643

\bibitem{Alekseeva_PhysicaScripta_PAS_1979}
Alekseeva O~K, Bykov V~N, Levdik V~A, Miron N~F and Shantarovich V~P 1979 {\em
  Physica Scripta\/} {\bf 20} 683--684

\bibitem{Hautojarvi_Vac_H_PRB_1985}
Hautojarvi P, Huomo H, Puska M and Vehanen A 1985 {\em Phys. Rev. B\/} {\bf 32}
  4326--4331

\bibitem{Romanenko_TFSRF_2012}
Romanenko A, Edwardson C and Coleman P Inner workings of the {120C} baking
  effect studied by positron annihilation, talk at {TFSRF'2012},
  www.jlab.org/indico/conferenceDisplay.py?confId=22

\end{thebibliography}

\providecommand{\newblock}{}

\end{document}